\documentclass[prc,floatfix,twocolumn,showpacs,superscriptaddress,amssymb,amsmath]{revtex4}

\usepackage{amssymb}
\usepackage{graphicx}
\usepackage{bm}

\begin{document}
\title{Gamma-Ray Bursts Black hole accretion disks as a site for the
  $\nu p$-process}

\author{L.-T.~Kizivat}
\affiliation{GSI Helmholtzzentrum f\"ur Schwerionenforschung,
  Planckstr. 1, 64291 Darmstadt, Germany}
\affiliation{Institut f{\"u}r Kernphysik, Technische Universit{\"a}t
  Darmstadt, 64289 Darmstadt, Germany} 
\author{G.~Mart\'inez-Pinedo} 
\affiliation{GSI Helmholtzzentrum f\"ur Schwerionenforschung,
  Planckstr. 1, 64291 Darmstadt, Germany} 
\author{K.~Langanke} 
\affiliation{GSI Helmholtzzentrum f\"ur Schwerionenforschung,
  Planckstr. 1, 64291 Darmstadt, Germany} 
\affiliation{Institut f{\"u}r Kernphysik, Technische Universit{\"a}t
  Darmstadt, 64289 Darmstadt, Germany} 
\affiliation{Frankfurt Institute for Advanced Studies,
  Ruth-Moufang-Str. 1, 60438 Frankfurt, Germany} 
\author{R.~Surman}
\affiliation{Department of Physics and Astronomy, Union College,
  Schenectady, NY 12308} 
\author{G.~C.~McLaughlin} 
\affiliation{Department of Physics, NC State University, Raleigh, NC
  27695-8202} 
\date{\today}

\begin{abstract}
  We study proton rich nucleosynthesis in windlike outflows from
  gamma-ray bursts accretion disks with the aim to determine if such
  outflows are a site of the $\nu p$-process.  The efficacy of this
  $\nu p$-process depends on thermodynamic and hydrodynamic
  factors. We discuss the importance of the entropy of the material,
  the outflow rate, the initial ejection point and accretion rate of
  the disk. In some cases the $\nu p$-process pushes the
  nucleosynthesis out to $A\sim 100$ and produces light
  p-nuclei. However, even when these nuclei are not produced, neutrino
  induced interactions can significantly alter the abundance pattern
  and cannot be neglected.
\end{abstract}

\maketitle
\section{Motivation and Methods\label{sec:level1}}

Burbidge, Burbidge, Hoyle and Fowler \cite{Burbidge.Burbidge.ea:1957}
and independently Cameron \cite{Cameron:1957} have proposed three
major nucleosynthesis processes - the r-, s-, and p-process - to make
nuclei heavier than iron in stars.  Since this pioneering work
impressive progress has been achieved in the understanding of these
processes \cite{Wallerstein.Iben.ea:1997}. Nevertheless important
questions still remain unanswered like the definite astrophysical site
of the r-process~\cite{arnould.goriely.takahashi:2007} and the
mechanism responsible for the production of light
p-nuclei~\cite{Arnould.Goriely:2003}.

Neutrino-driven winds from core-collapse supernovae are considered as
a site for the production of light p-nuclei. In this scenario, light
p-nuclei can be produced by different mechanisms. Slightly neutron
rich outflows with electron fractions $Y_e \approx 0.48$ result in the
production of light p-nuclei with $A\leq92$, i.e. $^{74}$Se, $^{78}$Kr
$^{84}$Sr, and $^{92}$Mo~\cite{Hoffman.Woosley.ea:1996}. The
production of some of these nuclei and in particular $^{92}$Mo is
enhanced by neutrino captures on nuclei with $N\approx
50$~\cite{Fuller.Meyer:1995}. Light-p nuclei, including $^{94}$Mo and
$^{96,98}$Ru, can be produced in proton-rich outflows via the $\nu
p$-process~\cite{Froehlich.Martinez-Pinedo.ea:2006,Pruet.Hoffman.ea:2006,Wanajo:2006}. It
is this latter mechanism that we focus on in this paper. It operates
in proton-rich environments with high neutrino and antineutrino fluxes
as are found in the early ejected matter in core-collapse supernovae.
Here the matter is ejected from the surface of the nascent neutron
star as free nucleons.  The competition of neutrino captures on
neutrons and antineutrino captures on protons drives the matter
proton-rich as both neutrino types have rather similar luminosities
and the average antineutrino energy is not large enough compared to
the neutrino energy to compensate for the difference in reaction
Q-values. Upon reaching cooler regions, i.e.  with increasing distance
from the neutron star surface, the nucleons assemble in nuclei and,
without further neutrino reactions, the proton-rich matter freezes out
with a significant production of $N=Z$ nuclei like $^{56}$Ni and
$^{64}$Ge and some free protons left. However, antineutrino captures
on these protons ensure a significant presence of free neutrons which
can be captured on the $N=Z$ nuclei via $(n,p)$ and $(n,\gamma)$
reactions allowing for matter flow beyond $^{56}$Ni and $^{64}$Ge
which otherwise with their long halflives against proton capture and
beta decay could not be overcome during the dynamical timescale of
supernova nucleosynthesis (a few seconds).  Different studies of the
$\nu
p$-process~\cite{Froehlich.Martinez-Pinedo.ea:2006,Pruet.Hoffman.ea:2006,Wanajo:2006}
have shown that it can produce light p-nuclides in abundances which
might be sufficient to explain their observed solar values.  It is
expected that the ejection of some proton-rich matter in the presence
of strong antineutrino fluxes is a general feature of core-collapse
supernovae making the occurrence of the $\nu p$-process a general
supernova phenomenon.

Recent studies indicate that quite similar physical conditions, like
those which give rise to the $\nu p$-process in supernovae, also occur
in the windlike outflows from the accretion disk surrounding a black
hole featuring a gamma-ray burst.  Considerable observational evidence
now connects long duration gamma ray bursts with the collapse of
massive stars.  McFadyen and Woosley \cite{Macfadyen.Woosley:1999}
pioneered a currently widely used model for these events: a collapse
of massive star with $M > 25$~M$_\odot$ (M$_\odot$ denoting our sun's
mass) fails to produce a standard core collapse supernovae with a
neutron star at the center, and instead produces an accretion disk
surrounding a black hole. Here again, the temperature in the disk is
sufficiently high so that matter is ejected as free
nucleons. Furthermore, the ejection occurs in the presence of quite
sizable neutrino fluxes \cite{Surman.Mclaughlin:2004} Depending on the
matter accretion rate of the black hole disk, neutrinos and
antineutrinos can get trapped defining neutrino surfaces from which
they decouple from the disk
matter~\cite{Surman.Mclaughlin:2004,Mclaughlin.Surman:2007}.  It is
found that disks with moderate accretion rates (around 1~M$_\odot$
s$^{-1}$) have neutrino and antineutrino fluxes which drive the
ejected matter proton-rich making the occurrence of the $\nu
p$-process an exciting possibility
\cite{Surman.Mclaughlin:2005}. Lacking detailed (magneto)hydrodynamic
models of the ejection, the amount of proton-rich ejecta can be
estimated to reach up to 0.01~M$_\odot$ according to the analytical
outflow model of ref.~\cite{Metzger.Piro.Quataert:2008} and typical
disk parameters. A broad survey of nucleosynthesis from GRB accretion
disks indicated that some p-process elements may be formed this way
\cite{Surman.Mclaughlin.Hix:2006}, but a study of the $\nu p$-process
mechanism has not been undertaken.  It is the aim of this paper to
study this intriguing possibility.  Our study is based on the
temperature, density, electron-to-nucleon ratios profiles and
dynamical expansion times of the windlike outflows as described by
McLaughlin and Surman
\cite{Surman.Mclaughlin:2004,Surman.Mclaughlin.Hix:2006}. The
respective wind trajectories are then coupled to an extensive nuclear
network.

\section{Model\label{Model}}

The disks studied in this work are all based on the model of
Neutrino-Dominated-Accretion-Flows,
``NDAF's''~\cite{Popham.Woosley.Fryer:1999}.  Two NDAF-based disk
models are used and results compared. The models labeled A1--A6 are
from a disk model by T.  DiMatteo, R. Perna and R. Narayan
\cite{Di.Perna.Narayan:2002}; this model was the first to incorporate
the effects of neutrino trapping.  The models labeled B1-B5 are from a
fully relativistic disk model by W.-X. Chen and A.M. Beloborodov
\cite{Chen.Beloborodov:2007} that additionally incorporates improved
microphysics, including the influence of electron degeneracy and an
evolving neutron-to-proton ratio.  For a detailed description of
neutrino-cooled accretion disks, see \cite{Di.Perna.Narayan:2002},
\cite{Popham.Woosley.Fryer:1999}, \cite{Chen.Beloborodov:2007}.
Radial profiles for the disk density $\rho$, temperature $T$, disk
scale height $H$, nuclear composition and neutrino and antineutrino
luminosity $L_{\nu,\bar\nu}$ are found to depend strongly on the
accretion disk parameters, i.e., the disk accretion rate $\dot{M}$,
viscosity parameter $\alpha$, black hole spin $a$ and accretion disk
mass $M$. The matter in the disk has to decrease its internal energy
to be accreted into the black hole. This cooling of the disk proceeds
either through advection or neutrino
emission~\cite{Chen.Beloborodov:2007}.  Cooling through neutrino
emission only becomes important in regions where the temperature and
density are sufficiently high for the below reactions to occur with
significant rates:
\begin{subequations}
\label{eq:rates}
\begin{eqnarray} 
  e^- + e^+ &\rightleftarrows &\nu_e + \bar{\nu_e} \label{eq:rates1}\\
  e^- + p &\rightleftarrows &\nu_e + n \label{eq:rates2}\\
  e^+ + n &\rightleftarrows & \bar{\nu_e} + p \label{eq:rates3}
\end{eqnarray}  
\end{subequations}

These neutrinos emitted from the disk interact with material ejected
from the disk, influencing the subsequent nucleosynthesis in the
outflow.  The neutrino and antineutrino fluxes above the disk are
calculated as in Surman and McLaughlin \cite{Surman.Mclaughlin:2004}.
Where the disk is optically thin to neutrinos and antineutrinos, the
emitted neutrino fluxes are determined from the rates of the above
reactions.  In the inner regions of the disk, first the neutrinos and
then the antineutrinos become trapped.  Here we determine the surfaces
at which electron neutrinos and antineutrinos decouple vertically from
the accretion disk and then use the local disk temperatures at these
places to adjust the temperatures of the neutrino and antineutrino
fluxes leaving the disk.

The evolutionary path of mass elements in an outflow is given in a
trajectory. The initial thermodynamic parameters of the matter in the
outflow are set by the characteristics of the disk, i.e. vary
according to the disk model used. Matter in the outflows leaves the
disk in perpendicular direction to the disk. Close to the disk a
treatment in cylindrical symmetry is convenient. With increasing
distance to the disk a transition from cylindrical to spherical
symmetry is appropriate.

The present calculations are based on the simulations of the windlike
outflows of Surman and McLaughlin \cite{Surman.Mclaughlin:2005}.
Using the method outlined in \cite{Surman.Mclaughlin:2005} we
construct parametrized trajectories which describe the ejection of
mass elements from the disk.  Due to the rather high temperatures
the matter is originally ejected as free neutrons and protons and the
proton-to-neutron ratio is set by weak interactions due to the strong
neutrino and antineutrino fluxes which occur rather close to the
neutrino surfaces. Upon reaching cooler regions at distances
noticeably further out than these surfaces nucleons can reassemble to
nuclei.  This nucleosynthesis process in the matter outflow we
describe by an extensive nucleosynthesis network subject to the
thermodynamical conditions of the outflowing matter.

\begin{figure}
  \centering
  \includegraphics[width=0.85\columnwidth]{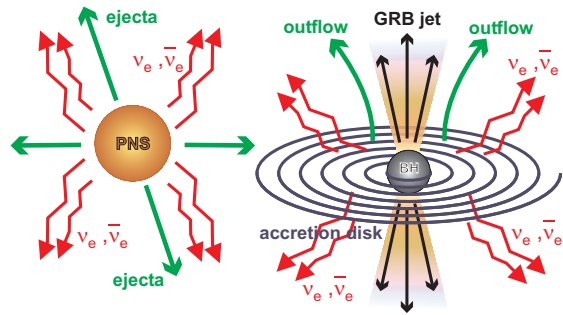}
  \caption{(Color online) Scheme of the geometry in core collapse
    supernovae and GRB accretion disks. ``PNS'' is the nascent
    proto-neutron star. Neutrinos and matter in the early ejecta are
    emitted radially. ``BH'' denotes the black hole, the central
    engine in a GRB accretion disk. The outflows leave the disk
    vertically.\label{fig:outflows}}
\end{figure}

We assume the outflows to be adiabatic defined by the entropy per
baryon $S$ in units of k$_B$. A second crucial parameter is the
velocity of the ejected matter as a function of distance $R$ from the
black hole in the center of the disk. This velocity $u$ is
parameterized as
\begin{equation} \label{eq:velocity} \mid u \mid = v_\infty \left(1 -
    \frac{R_0}{R}\right)^\beta.
\end{equation}
where $R_0$ is the radius at which the outflow leaves the accretion
disk (see Figure~\ref{fig:outflows}), $v_\infty$ is the asymptotic
velocity of the matter (assumed to be $ 3 \times 10^4$ km/s) and the
parameter $\beta$ determines the acceleration of the outflow and
therefore is referred to as ``acceleration
parameter''~\cite{Surman.Mclaughlin:2005}.  The parameter $\beta$ is
important as it defines the time the outflow is subjected to weak
interactions in the strong neutrino fluxes.  Large values of $\beta$
correspond to slowly accelerating outflows with rather long neutrino
interaction times, while small $\beta$ values define outflows with
fast accelerations and hence less time for neutrino interactions.

As pointed out in \cite{Surman.Mclaughlin:2005} the electron fraction
$Y_e$ of the ejected matter depends also critically on the rate by
which the black hole accretes matter to the disk.  Typical values are
in the range 0.01--10 M$_\odot$ s$^{-1}$~\cite{Popham.Woosley.Fryer:1999}.
Disks with smaller accretion rates of order 0.1~M$_\odot$ s$^{-1}$ and
1~M$_\odot$ s$^{-1}$, as explored in our current work, are found to have only
small regions of trapped neutrinos in the disk and even smaller ones
for antineutrinos. Under such conditions there are relatively few
antineutrinos interacting with the matter outflow, while neutrino
captures on neutrons are still sizable. For such low accretion rates
the weak interaction will drive the outflowing matter proton-rich
\cite{Surman.Mclaughlin:2005} with free protons available making it a
tempting site for $\nu p$-process nucleosynthesis. It is exactly these
conditions which we will study in the following.

We note that for even smaller accretion rates neither neutrino nor
antineutrino trapping occurs and the free proton fraction in the
ejected matter is not significant. Such an environment is not
favorable for $\nu p$-process nucleosynthesis.

Our nucleosynthesis network considers 3347 nuclei covering the nuclear
chart from protons and neutrons to ${}^{211}\textrm{Eu}$. The
appropriate reaction rates among these nuclei mediated by the strong
and electromagnetic interaction are taken from the reaction library
\cite{Rauscher.Thielemann:2000}. Weak-interaction rates for nuclei are
adopted from Zinner~\cite{Zinner:2007}. However, we find that only
neutrino reactions with free protons and neutrons have influence on
the nucleosynthesis process. The network has been numerically solved
as outlined in~\cite{Hix.Meyer:2006,Hix.Thielemann:1999b}.

\section{Results}

As outlined above we expect that the nucleosynthesis in the accretion
disk outflows depends on the parameter values for the mass accretion
rate $\dot{M}$, the acceleration parameter $\beta$, the radius $R_0$
at which the matter decouples from the disk and the entropy in the
outflow $S$. Furthermore, the proton-to-neutron ratio of the ejected
matter is crucial for the nucleosynthesis.  As this ratio is set by
weak interaction processes close to the neutrino surfaces it is
expected to be different for the neutrino-dominated accretion disk
models of DiMatteo et al. \cite{Di.Perna.Narayan:2002} and of Chen and
Beloborodov \cite{Chen.Beloborodov:2007} which predict quite
noticeable differences in the relative neutrino and anti-neutrino
fluxes and spectra.  Hence we have performed a set of nucleosythesis
calculations for both models varying the parameters $\dot{M}$,
$\beta$, $R_0$ and $S$.  The chosen parameter values for the various
models which we studied in details are defined in Table
\ref{tab:model}, where the labels 'A' and 'B' refer to the disk models
of \cite{Di.Perna.Narayan:2002} and \cite{Chen.Beloborodov:2007},
respectively.

\begin{table}
  \begin{tabular}{ccccc}
    \hline\hline
    model & $\dot{M}$ & $R_0$  & $\beta$ & $S$ \\
    & (M$_\odot$ s$^{-1}$)  & (km) &   &  ($k_B$) \\ \hline
    A1 \& B1   &  1     & 100 & 2.5 & 30 \\
    A2 \& B2   &  1     & 100 & 0.8 & 30 \\
    A3 \& B3   &  1     & 100 & 2.5 & 15 \\
    A4 \& B4   &  1     & 100 & 2.5 & 50 \\
    A5            &  1     & 250 & 2.5 & 50 \\
    B5            &  1     &  50 & 2.5 & 50 \\
    A6	       &  0.1   & 100 & 2.5 & 30 \\
    \hline\hline
  \end{tabular}
  \caption{Parameter values for the various studied
    models.\label{tab:model}} 
\end{table}

\begin{figure}
  \centering
  \includegraphics[width=\linewidth]{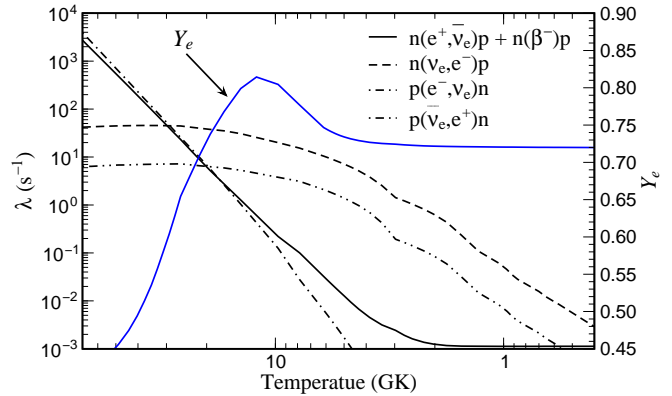}
  \caption{(Color online) Model A1: Evolution of the capture rates for positrons
    (solid line) and neutrinos (dashed) on neutrons and for electrons
    (dashed-dotted) and antineutrinos (dashed-dotted-dotted) on
    protons as a function of the matter temperature.  Notice that the
    rate for positron capture on neutrons includes also the neutron
    beta decay rate. The evolution of the electron fraction, $Y_e$ is
    also shown (right y-scale). Temperature is measured from the
    moment matter decouples from the disk where it is still strongly
    neutron-rich.\label{fig:captures_A1}}
\end{figure}

\begin{figure}
\includegraphics[width=\linewidth]{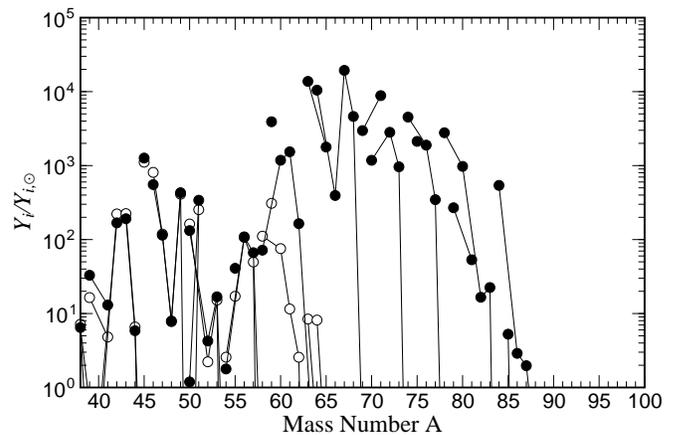}
\caption{Final isotopic abundances (solid circles) for the
  nucleosynthesis calculations in disk model A1 relative to solar
  abundances~\cite{Lodders:2003}.  The empty circles show the
  abundances obtained with the same nucleosynthesis network, however,
  switching off neutrino and anti-neutrino capture reactions once the
  freeze-out value for $Y_e$ is reached.\label{fig:A1}}
\end{figure}

We start with nucleosynthesis studies for the outflows from the disk
model of DiMatteo \emph{et al.} \cite{Di.Perna.Narayan:2002}.  The
Model A1 with the modest accretion rate $\dot{M} = 1$~M$_\odot$ s$^{-1}$ and
relatively slow outflow velocities corresponds to conditions for which
one expects the outflows to be proton-rich as at the radius $R_0$ at
which matter is expelled from the disk (100 km) the disk is dense and
hot enough for neutrino producing reactions (Eq.~\ref{eq:rates}) to
occur with 
significant rates.  Figure~\ref{fig:captures_A1}, shows the time evolution
of the electron fraction $Y_e$ and the capture rates of positrons and
neutrinos on neutrons and electrons and antineutrinos on protons. Time
is measured from the moment matter decouples from the disk where it is
still strongly neutron-rich and consists of free protons and
neutrons. Their ratio, however, is changed due to fast
electron-positron captures and interactions with the large neutrino
and antineutrino fluxes. Importantly the neutrino fluxes are about one
order of magnitude larger than those for antineutrinos, while the
average energy of antineutrinos is around 2~MeV larger than the one of
neutrinos. As a consequence neutrino captures on neutrons dominate and
drive the composition proton-rich and a peak value of $Y_e \sim 0.8$
is reached.  The increase in $Y_e$ is stopped, once alpha particles
form which use up all available neutrons. Due to the high particle
thresholds of $^4$He neutrons are then protected against neutrino
interactions, while the remaining free protons are still subject to
strong antineutrino fluxes. Continuing antineutrino captures on
protons supply more neutrons, which, in an ``inverse'' alpha effect
\cite{Fuller.Meyer:1995,Mclaughlin.Fuller.Wilson:1996,
Meyer.Mclaughlin.Fuller:1998} 
are combined with additional protons into more $^4$He. As a
consequence of these antineutrino captures the $Y_e$ value
decreases. At $T \sim 3$~GK the weak interaction rates have decreased
sufficiently as the neutrino fluxes decrease with distance and the
matter composition freezes out with a $Y_e$ value around
0.72. Starting with the triple-alpha reaction a network of reactions
(called alpha process
\cite{Woosley.Hoffman:1992,Witti.Janka.Takahashi:1994}) produces
heavier nuclei, which can become the seed for additional
nucleosynthesis, and reduces the abundance of $^4$He. For conditions
with $Y_e > 0.5$, the seeds consist mainly of $N=Z$ nuclei with
multi-alpha structures like $^{56}$Ni, $^{60}$Zn, $^{64}$Ge, with an
abundance of free protons $Y_p = 2
Y_e-1$~\cite{Pruet.Hoffman.ea:2006}. The long halflives of these seed
nuclei against beta decay and proton capture would stop the
nucleosynthesis flow, if it were not for the continuous supply of free
neutrons produced by antineutrino captures on the protons. This is the
key element of the $\nu p$-process. Indeed a sequence of $(n,p)$ and
$(p,\gamma)$ reactions allows for synthesis of elements upto the mass
range $A \sim 80-100$ for the conditions of model A1, as can be seen
in Fig.~\ref{fig:A1}. As a comparison we also show the abundance
distribution obtained with the same nucleosynthesis network, however,
switching off the neutrino and, importantly, antineutrino capture
reactions once the freeze-out value of $Y_e$ is reached. Strikingly
the matter flow stops at the alpha seed nuclei with long halflives
(e.g. $^{56}$Ni, $^{60}$Zn..) as in this scenario no free neutrons are
available to carry the matter flow to larger mass numbers as it is
guaranteed in the $\nu p$-process due to the late-time neutrons
produced by antineutrino captures on protons.

\begin{figure}
\includegraphics[width=\linewidth]{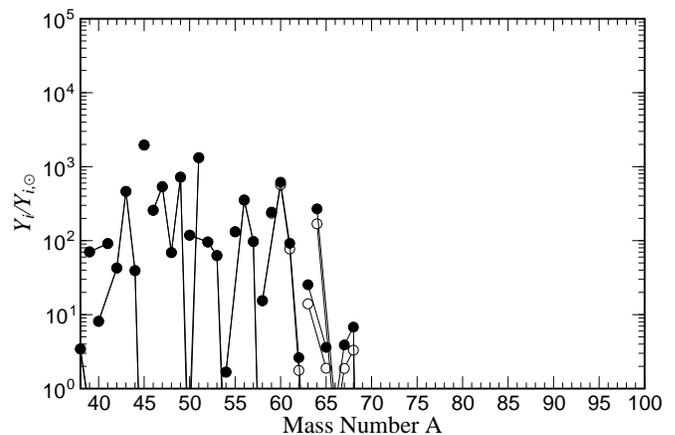}
\caption{Same as Fig. \ref{fig:A1}, but for model A2.\label{fig:A2}}
\end{figure}

We have seen that model A1 allows for a substantial $\nu p$-process to
occur.  The outflow model A2 is identical to model A1, except that it
has a faster acceleration (smaller $\beta$ value). Thus, although the
neutrino and antineutrino luminosities are the same as in model A1,
the ejected matter has shorter time to interact with neutrinos and
consequently the neutrino fluence is much smaller. While in model A1
the neutrino fluence in the temperature range 3--1~GK is $4\times
10^{38}$~cm$^{-2}$ (a value similar to the one used in $\nu p$-process
nucleosynthesis studies in supernova environments~\cite{Froehlich.Martinez-Pinedo.ea:2006,Pruet.Hoffman.ea:2006}) the
value for model A2 is only $10^{37}$~cm$^{-2}$. Consequently, model A2
is expected to show only a rather weak $\nu p$-process which is indeed
borne out by our nucleosynthesis calculation, as can be seen in
Fig.~\ref{fig:A2}.

\begin{figure}
\includegraphics[width=\linewidth]{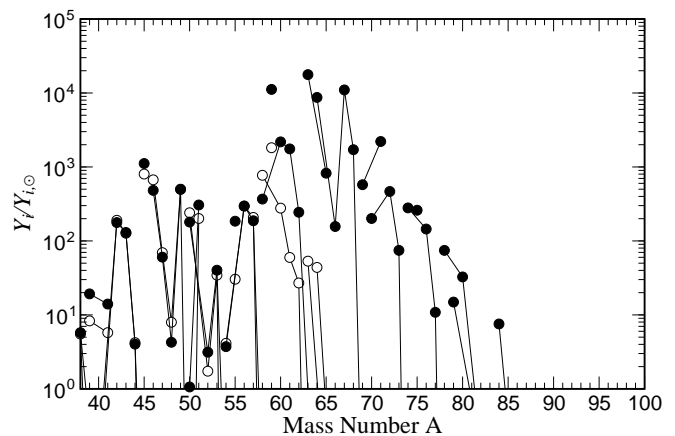}
\caption{Same as Fig. \ref{fig:A1}, but for model A3.\label{fig:A3}}
\end{figure}

\begin{figure}
\includegraphics[width=\linewidth]{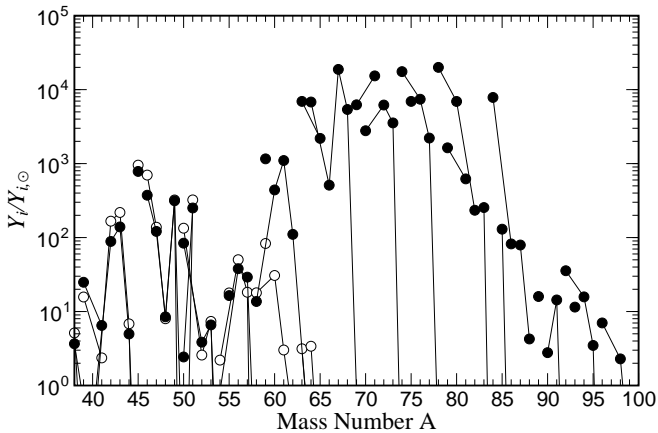}
\caption{Same as Fig. \ref{fig:A1}, but for model A4.\label{fig:A4}}
\end{figure}

Another important parameter, which affects the matter composition and
the subsequent nucleosynthesis, is the entropy $S$. In general the
larger the entropy, the larger the fraction of free protons which are
available per seed nucleus. To study the influence of entropy on the
final nucleosynthesis abundances we have repeated the parameter study
of model A1, however, replacing the entropy of the ejected matter by a
smaller value (15 $k_B$, model A3) and by a larger value (50 $k_B$,
model A4) than used in model A1 (30 $k_B$). We mention again that the
neutrino fluxes are the same in models A1, A3, and A4.  The fact that
for model A3 (15 $k_B$), with lower entropy, there are less free
protons available per seed nuclei translates into a smaller supply of
neutrons produced by antineutrino captures and a less pronounced $\nu
p$-process as observed for model A1. The behavior is opposite for
model A4 (50 $k_B$), where the larger entropy reults in more free
protons per seed nucleus, and consequently more free neutrons per
seed, than model A1. These observations are confirmed by the abundance
distributions shown in Figs.  \ref{fig:A3} and \ref{fig:A4}.  Model A4 (50
$k_B$) indeed shows a substantial amount of elements in the mass range
$A \sim 60-100$, while model A3 (15 $k_B$) produces only nuclides up to
mass $A \approx 80$ in a noticeably weaker $\nu p$-process.

\begin{figure}
\includegraphics[width=\linewidth]{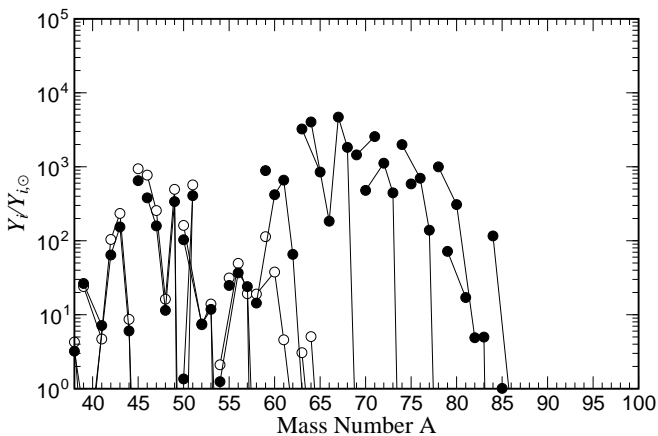}
\caption{Same as Fig. \ref{fig:A1}, but for model A5.\label{fig:A5}}
\end{figure}

As mentioned before, the disks have regions of different nuclear
composition, density and temperature.  The weak interaction rates in
the outflows depend on the neutrino and antineutrino fluxes
originating from the disk.  The flux magnitudes in turn reflect disk
conditions at certain radii. Thus the radius $R_0$, at which matter is
released from the disk, affects $\nu p$-process nucleosynthesis as
weak interaction rates in the outflow depend on the magnitude of the
neutrino and antineutrino fluxes. In disks with comparable accretion
rate at larger radii neutrinos will drive the composition less
proton-rich as due to disk composition the neutrino fluxes are
smaller. At smaller radii neutrinos will drive the composition more
proton-rich and neutrino fluxes are larger. 

To demonstrate this
quantitatively we have repeated the nucleosynthesis calculation of
model A4, however, replacing the radius at which matter is released to
a larger value of $R_0 = 250$ km (model A5).  As expected, a weaker
$\nu p$-process occurs in model A5 (Fig.~\ref{fig:A5}). 

\begin{figure}
  \centering
  \includegraphics[width=\linewidth]{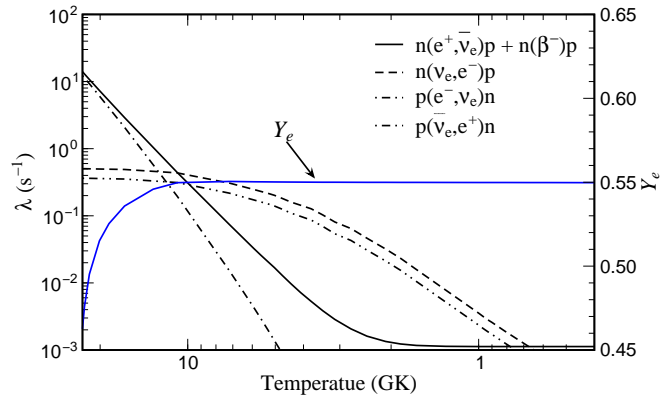}
  \caption{(Color online) Same as Fig.~\ref{fig:captures_A1}, but for model
    A6.\label{fig:captures_A6}}
\end{figure}

\begin{figure}
\includegraphics[width=\linewidth]{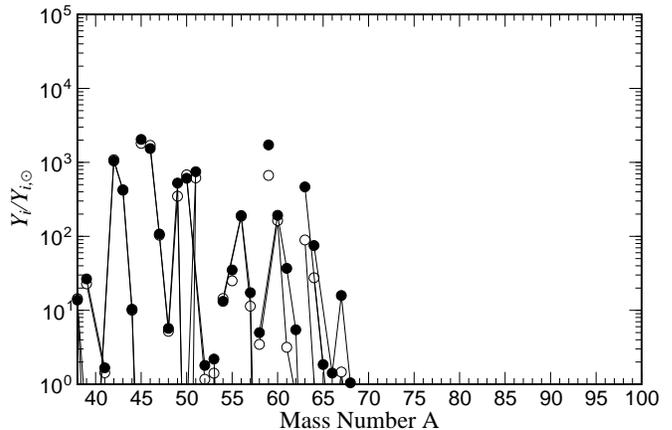}
\caption{Same as Fig. \ref{fig:A1}, but for model A6.\label{fig:A6}}
\end{figure}

Model A6 is the same as model A1, except that the matter accretion
rate is significantly lower, $\dot {M} = 0.1$ M$_\odot$ s$^{-1}$
rather than 1~M$_\odot$ s$^{-1}$ as assumed in all other models. As is
explained in~\cite{Surman.Mclaughlin:2005} this change in accretion
rate has significant effects on the (anti-)neutrino
surfaces. Consequently it also strongly affects the neutrino and
antineutrino fluxes which now are quite similar rather than dominated
by a much stronger neutrino than antineutrino flux as encountered in
the other models. The fluxes are in overall a factor of magnitude
smaller than in models with higher accretion rate. Although the
average antineutrino energy is slightly larger than the average
neutrino energy, it is not the neutrino capture reactions which drive
the composition slightly proton-rich in this model (see
Fig. \ref{fig:captures_A6}) but positron captures on free
neutrons. The rates for the discussed reactions (Eq.~\ref{eq:rates})
depend on the temperature and density of the disk. In fact, in model
A6 a sufficient amount of positrons is produced which drives reaction
(\ref{eq:rates3}) to the right and makes the ejected matter
proton-rich, reaching $Y_e$ values up to 0.55. As is explained for
model A1, once nuclei can be formed all neutrons are blocked in $^4$He
with some free protons left. The formation of $^4$He occurs before
weak freeze out.  However, this value for $Y_e$ is significantly
smaller than in model A1 ($Y_e \sim 0.72$). This fact together with
the smaller antineutrino fluxes results in a smaller production of
neutrons per seed nuclei and a rather weak $\nu p$-process, as is
demonstrated in Fig. \ref{fig:A6}.

We now switch to the nucleosynthesis studies for the outflows from the
disk model of Chen and Beloborodov~\cite{Chen.Beloborodov:2007}. This
model predicts neutrino fluxes and spectra which are quite distinct
from those of the disk model of DiMatteo \emph{et al.}
\cite{Di.Perna.Narayan:2002}. The differing treatments of relativity
and the microphysics of the two disk models results in markedly
different disk temperature and density profiles, particularly in the
innermost regions with the greatest neutrino emission.  In the
DiMatteo \emph{et al.} disk model, the temperature and density rise
more steeply with decreasing radius and the innermost regions are
significantly hotter and denser than in the Chen and Beloborodov disk
model, resulting in correspondingly higher neutrino emission.

\begin{figure}
\centering
\includegraphics[width=\linewidth]{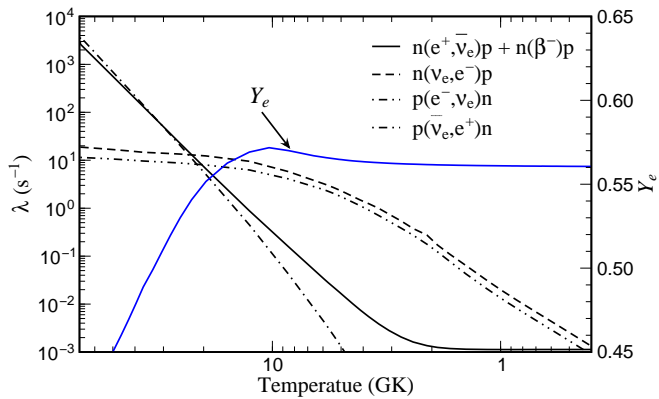}
\caption{(Color online) Same as Fig.~\ref{fig:captures_A1}, but for model
    B1. \label{fig:captures_B1}}
\end{figure}
\begin{figure}
\includegraphics[width=\linewidth]{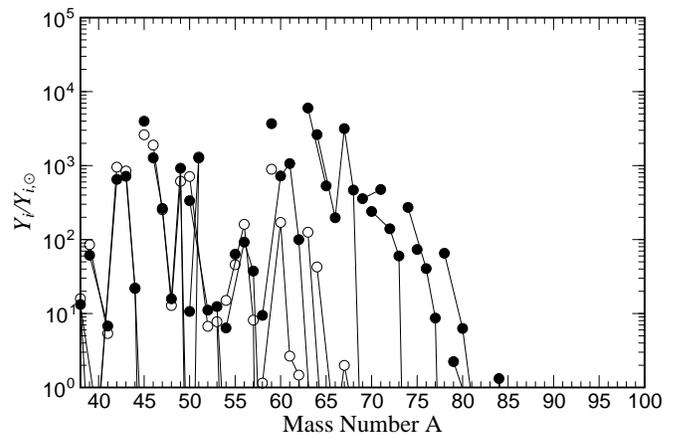}
\caption{Same as Fig.~\ref{fig:A1}, but for model
  B1.\label{fig:B1}} 
\end{figure}

To explore the impact of these differences on the nucleosynthesis we
have performed nucleosynthesis studies for the Chen and Beloborodov
model (called models B1-B4) using the same parameters for the mass
accretion rate, the acceleration parameter, the decoupling radius, and
the entropy as in models A1-A4 above (see Table~\ref{tab:model}).
Fig.~\ref{fig:captures_B1} is the equivalent to Fig. \ref{fig:captures_A1}
clearly showing impact of the different neutrino fluxes on the
proton-to-neutron ratio of the ejected matter. 
In contrast to model A1, in which the neutrino luminosity exceeds the
one for anti-neutrinos, by about an order of magnitude, the neutrino
and anti-neutrino fluxes are rather similar in model B1. As a
consequence also the neutrino and anti-neutrino capture rates on
neutrons and protons, respectively, differ significantly less than in
model A1. Moreover, neutrino and anti-neutrino capture rates are both
smaller than the rates for positron and electron captures during the
first second after matter has decoupled from the disk. This situation
is similar to model A6, and again it is the slight dominance of
positron over electron captures which drive the matter proton-rich at
first.  There is only a little time period in model B1 for which
neutrino and anti-neutrino captures dominate the inverse
weak-interaction processes. As the anti-neutrino capture rate is
slightly larger by a factor of order 2, the ejected matter is further
driven proton-rich during this short period.  The final
proton-to-neutron ratio $Y_e = 0.56$ is much smaller than in model A1
($Y_e = 0.72$).  We also observe an inverse alpha-effect in model B1,
but it is milder than in model A1.  Since $Y_e > 0.5$, the $\nu
p$-process can also operate in model B1. However, its effectiveness is
noticeably less than observed in model A1, as less free protons are
available with a reduced antineutrino flux
(see Fig. \ref{fig:B1}). While nuclides beyond $A=64$ are
being made, the $\nu p$-process stops in model B1 already around $A =
80$ and produces also smaller abundances of nuclides in the mass range
$A=60-80$ than model A1. 

\begin{figure}
  \includegraphics[width=\linewidth]{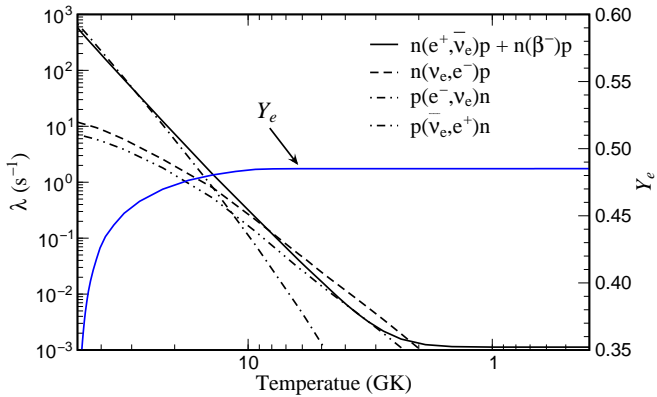}
  \caption{(Color online) Same as Fig. \ref{fig:captures_A1}, but for model
    B2.\label{fig:captures_B2}} 
\end{figure}

\begin{figure}
  \includegraphics[width=\linewidth]{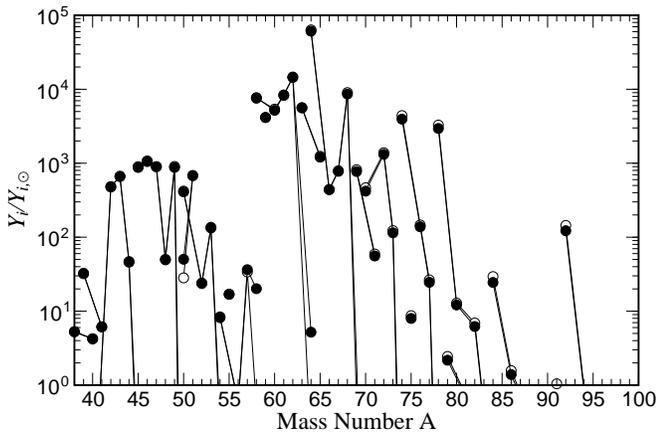}
  \caption{Same as Fig.\ref{fig:A1}, but for model B2.\label{fig:B2}}
\end{figure}

Model B2 is the same as model B1, except that it has a smaller value
for $\beta$ (faster acceleration). Hence weak-interaction processes
have less time to change the proton-to-neutron ratio. In fact, as can
be seen in Fig.~\ref{fig:captures_B2}, the final value for $Y_e$ is
already been reached at a time of 0.1 s after the matter decoupled
from the disk. More importantly for the nucleosynthesis, the value of
$Y_e$ stays below 0.5 in this model, i.e. the ejected matter is
neutron-rich.  We also observe that positron and electron captures
dominate over neutrino and anti-neutrino captures during the time
period in which the $Y_e$ value of the ejected matter is set to its
final value by weak interactions. With the matter being slightly
neutron-rich, obviously no $\nu p$-process occurs.  Fig. \ref{fig:B2} shows
that the abundance distributions are nearly identical with and without
consideration of neutrino and anti-neutrino reactions in the nuclear
network after the final $Y_e$ value has been reached.  These abundance
distributions resemble those of an $\alpha$-process for a given value
of $Y_e < 0.5$.  The distinct differences in the abundance
distributions of an alpha-rich freeze-out for proton-rich ($Y_e >
0.5$) and neutron-rich ($Y_e < 0.5$) matter is explained in
\cite{Seitenzahl.Timmes.ea:2008}.)  We note that such an
$\alpha$-process operates also in the neutrino-driven wind scenario
setting up the abundance distribution for the seed nuclei of an
subsequent r-process. However, in our model B2 the entropy is
significantly smaller than required for successful r-process
simulations in the neutrino-driven wind model.  Due to this low
entropy the neutron-to-seed ratio is too low in model B2 to allow for
any r-process to occur. We note that the nuclear network considered in
our nucleosynthesis studies is large enough to indicate potential
r-process nucleosynthesis beyond the $\alpha$-process.

\begin{figure}
\includegraphics[width=\linewidth]{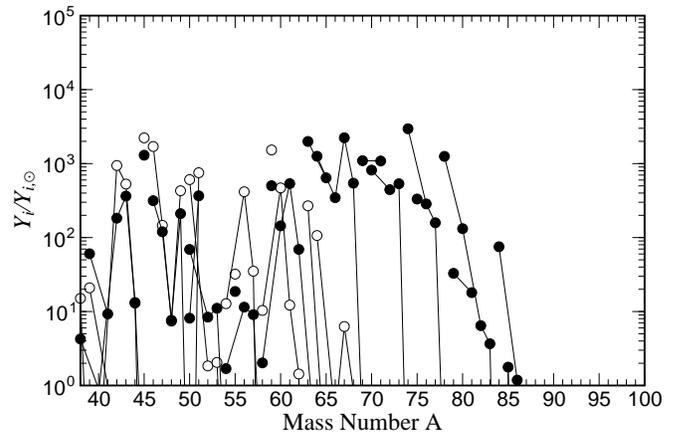}
\caption{Same as Fig. \ref{fig:A1}, but for model B3.\label{fig:B3}}
\end{figure}

\begin{figure}
\includegraphics[width=\linewidth]{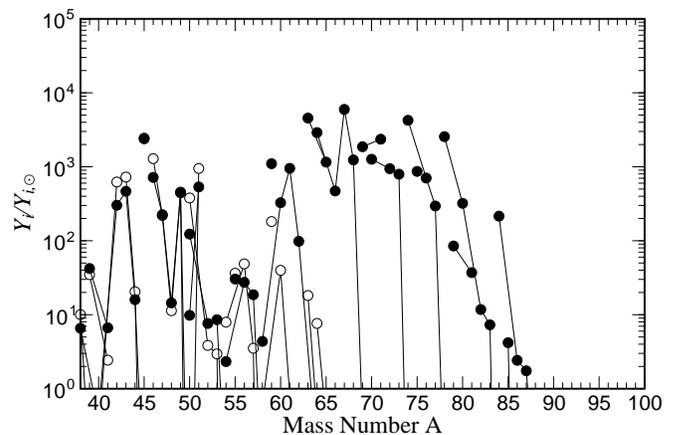}
\caption{Same as Fig. \ref{fig:A1}, but for model B4.\label{fig:B4}}
\end{figure}

Models B3 and B4 are the same as model B1, however, assuming a smaller
or larger value for the entropy ($S=15 k_B$, model B3 and $S=50 k_B$,
model B4, respectively).  As explained above, we expect a more
pronounced $\nu$p-process to occur with growing entropy, as a larger
amount of free protons (and hence neutrons after anti-neutrino
captures) is available once heavy nuclei are being formed in the
ejected matter. In fact, this is born out by our nucleosynthesis
studies for models B3 and B4 (see Figs.  \ref{fig:B3} and \ref{fig:B4}), if
compared to model B1.  It is also interesting to compare the abundance
results of models B3 and B4 to the equivalent studies of the DiMatteo
{\it et al.} disk model (models A3 and A4).  In both cases, the
stronger $\nu$p-process nucleosynthesis occurs for the DiMatteo {\it
  et al.}  disk models. The reason is the same as discussed in the
comparison of models A1 and B1.  The reduced enhacement of neutrino
over anti-neutrino fluxes in the Chen and Beloborodov models
translates into smaller $Y_e$ values than in the corresponding
DiMatteo {\it et al.} models.  The consequence are smaller
proton-to-seed (or neutron-to-seed) ratios available once the
$\nu$p-process gets operable.

\begin{figure}
\includegraphics[width=\linewidth]{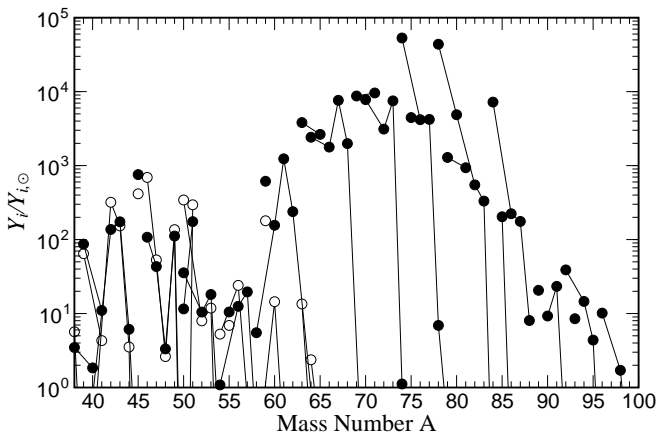}
\caption{Same as Fig. \ref{fig:A1}, but for model B5.\label{fig:B5}}
\end{figure}

Finally, in model B5 we explore the effect of reducing the radius
$R_0$ at which matter is released from the disk. This model is
identical to B4, however, it uses a value for $R_0$ of 50~km rather
than 100~km. As discussed above, this change leads to an increase in
the neutrino fluxes that will drive the composition more proton-rich
and to a stronger $\nu p$-process. This expectation is indeed borne out
in our calculations (see fig.~\ref{fig:B5}), which shows noticeable
strong production of elements with $A>75$. In contrast to B4 (see
fig.~\ref{fig:B4}), model B5 synthesizes also elements in the mass range
$A=90$--100.

\section{Conclusions}

Gamma-ray bursts are likely connected to an accretion disk surrounding
a black hole.  It has been found that the matter ejected from this
disk can exhibit a rather rich range of nucleosynthesis 
\cite{Surman.Mclaughlin.Hix:2006}
dependent on the matter accretion rate of the disk, the velocity and
entropy of the outflow as well as the radius at which the matter is
released. The composition of the outflow can be either neutron-rich,
giving range to r-process like nucleosynthesis \cite{Surman.Mclaughlin:2005}, or
proton-rich. In this manuscript we have chosen the parameters
characterizing the outflow such to focus on proton-rich
nucleosynthesis and to investigate whether a $\nu p$-process can occur
under reasonable conditions in the windlike outflows from a GRB
accretion disk. This is indeed confirmed by our extensive nuclear
network simulations. We find that it is essential to include neutrino
capture interactions, particularly anti-neutrino capture on protons,
when determining the nucleosynthetic outcomes for these environments.

In our present calculations, including disk models with accretion
rates from 0.1~M$_\odot$ s$^{-1}$ and 1~M$_\odot$ s$^{-1}$, we find
that disk models with accretion rates of order 1~M$_\odot$~s$^{-1}$
are particularly favorable for $\nu p$-process nucleosynthesis as
these models exhibit significantly larger neutrino fluxes than
antineutrino fluxes driving the matter proton-rich by a competition of
neutrino captures on neutrons and antineutrino captures on
protons. Our network calculations show that for reasonable parameters
describing velocity, entropy and ejection radius of the outflow a $\nu
p$-process can occur which produces substantial abundances of nuclides
up to the $A \sim 100$ mass range. The effectiveness of the production
in the mass range $A \sim 60$--100 depends somewhat on the entropy of
the outflow; where larger entropies support a larger amount of free
protons per seed nucleus and hence give a stronger $\nu p$-process as,
by antineutrino capture, the free protons supply the neutrons for $\nu
p$-process nucleosynthesis.

A crucial parameter for successful $\nu p$-process nucleosynthesis is
the acceleration parameter $\beta$ of the ejected matter as it defines
the time matter is subjected to the neutrino interactions. Obviously
the bigger the acceleration parameter, the more neutrino captures
drive the matter proton-rich. Indeed by varying the acceleration
parameter by a factor 3 we have demonstrated that the nucleosynthesis
outcome can change from the occurrence of a strong $\nu p$-process to
no $\nu p$-process nucleosynthesis.  Slightly weaker (stronger) $\nu
p$-process nucleosynthesis is also observed if the matter is released
at larger (smaller) radii from the disk due to reduced (enhanced) neutrino
capture rates.

It is also worth mentioning that our calculations show the occurrence
of a process which in analogy to the alpha effect in r-process
nucleosynthesis~\cite{Fuller.Meyer:1995,Mclaughlin.Fuller.Wilson:1996,
Meyer.Mclaughlin.Fuller:1998} 
we like to call ``inverse'' alpha effect.
If the formation of $^4$He, which in proton-rich environment locks up
all neutrons, occurs already before weak interaction freeze out, then
continuous antineutrino capture on protons supply more neutrons, which
combine with protons to build more $^4$He. This inverse alpha effect
reduces the proton-to-neutron ratio in proton-rich environments (while
in neutron-rich conditions it increases $Y_e$).  The inverse alpha
effect is particularly important for outflows with rather slow
acceleration parameters.

Accretion disk models are currently under development by several
groups, so
we have compared nucleosynthesis for the outflows of two different
accretion-disk models. While we find the same basic conclusion: that
the $\nu$-p process occurs, the details depend on the particular
accretion disk model used.  The model of DiMatteo \emph{et al.}
predicts a noticeable excess of neutrino flux over anti-neutrino
flux. This difference is much milder in the disk model of Chen and
Beloborodov. As a consequence of these differences in relative fluxes,
the ejected matter in the disk model of
Ref. \cite{Di.Perna.Narayan:2002} has always larger $Y_e$ values than
found for the model of \cite{Chen.Beloborodov:2007}.  Obviously the
occurence of an $\nu$p-process is always more pronounced for the
former model if nucleosynthesis abundances are compared for the same
values of mass accretion rate, matter acceleration, decoupling radies,
and entropy. With the models A2 and B2 we have presented examples,
where one disk model \cite{Di.Perna.Narayan:2002} predicts the
occurence of an $\nu$p-process, while, for the same parameter values,
for the other \cite{Chen.Beloborodov:2007} matter is ejected with a
$Y_e$ value less than 0.5 translating into an abundance distribution
known from $\alpha$-rich freeze-out for neutron-rich matter.

In summary, within our parameter studies we have shown that outflow
from accretion disks surrounding a black hole can be the site of $\nu
p$-process nucleosynthesis.  More quantitative studies of this
exciting perspective have to wait until complete hydrodynamical
simulations for the outflows from the accretion disks become
available.

\begin{acknowledgments}
  We thank B. D. Metzger for useful discussions. The work of
  LTK, GMP and KL was partially supported by the Helmholtz Alliance
  \emph{Cosmic Matter in the Laboratory} and the Deutsche
  Forschungsgemeinschaft through contract SFB 634. This work was
  partially supported by the Department of Energy under contract
  DE-FG02-02ER41216 (GCM) and under contract DE-FG02-05ER41398 (RS).
\end{acknowledgments}


\end{document}